\documentclass{article}
\usepackage{amsmath, amssymb, amsfonts}
\title{Anisotropic fluid in a time dependent conformally flat spacetime} 
\author{Hristu Culetu\\ Ovidius University, Dept.of Physics and Electronics, \\B-dul Mamaia 124, 900527 Constanta, Romania \\ email : hculetu@yahoo.com}
\begin{document}
\numberwithin{equation}{section}
\pagenumbering{arabic}
\maketitle
\begin{abstract}
The special conformal transformation (composed by inversion - translation - inversion) is used to generate a time dependent conformally flat spacetime. In order to be an exact solution of Einstein's equations, we need as a source a stress tensor corresponding to an anisotropic fluid with negative regular energy density and positive pressures. For the static approximation, the generators of the infinitesimal transformation resemble those recently obtained by Majhi and Padmanabhan for the coordinate transformation leading to the ''near horizon'' Schwarzschild metric in Kruskal coordinates. The static approximation corresponds to an energy momentum tensor of $\Lambda$ - form, the ''cosmological constant'' $\Lambda$ being proportional to the acceleration squared.

\end{abstract}
\newcommand{\fv}{\boldsymbol{f}}
\newcommand{\tv}{\boldsymbol{t}}
\newcommand{\gv}{\boldsymbol{g}}
\newcommand{\OV}{\boldsymbol{O}}
\newcommand{\wv}{\boldsymbol{w}}
\newcommand{\WV}{\boldsymbol{W}}
\newcommand{\NV}{\boldsymbol{N}}
\newcommand{\hv}{\boldsymbol{h}}
\newcommand{\yv}{\boldsymbol{y}}
\newcommand{\RE}{\textrm{Re}}
\newcommand{\IM}{\textrm{Im}}
\newcommand{\rot}{\textrm{rot}}
\newcommand{\dv}{\boldsymbol{d}}
\newcommand{\grad}{\textrm{grad}}
\newcommand{\Tr}{\textrm{Tr}}
\newcommand{\ua}{\uparrow}
\newcommand{\da}{\downarrow}
\newcommand{\ct}{\textrm{const}}
\newcommand{\xv}{\boldsymbol{x}}
\newcommand{\mv}{\boldsymbol{m}}
\newcommand{\rv}{\boldsymbol{r}}
\newcommand{\kv}{\boldsymbol{k}}
\newcommand{\VE}{\boldsymbol{V}}
\newcommand{\sv}{\boldsymbol{s}}
\newcommand{\RV}{\boldsymbol{R}}
\newcommand{\pv}{\boldsymbol{p}}
\newcommand{\PV}{\boldsymbol{P}}
\newcommand{\EV}{\boldsymbol{E}}
\newcommand{\DV}{\boldsymbol{D}}
\newcommand{\BV}{\boldsymbol{B}}
\newcommand{\HV}{\boldsymbol{H}}
\newcommand{\MV}{\boldsymbol{M}}
\newcommand{\be}{\begin{equation}}
\newcommand{\ee}{\end{equation}}
\newcommand{\ba}{\begin{eqnarray}}
\newcommand{\ea}{\end{eqnarray}}
\newcommand{\bq}{\begin{eqnarray*}}
\newcommand{\eq}{\end{eqnarray*}}
\newcommand{\pa}{\partial}
\newcommand{\f}{\frac}
\newcommand{\FV}{\boldsymbol{F}}
\newcommand{\ve}{\boldsymbol{v}}
\newcommand{\AV}{\boldsymbol{A}}
\newcommand{\jv}{\boldsymbol{j}}
\newcommand{\LV}{\boldsymbol{L}}
\newcommand{\SV}{\boldsymbol{S}}
\newcommand{\av}{\boldsymbol{a}}
\newcommand{\qv}{\boldsymbol{q}}
\newcommand{\QV}{\boldsymbol{Q}}
\newcommand{\ev}{\boldsymbol{e}}
\newcommand{\uv}{\boldsymbol{u}}
\newcommand{\KV}{\boldsymbol{K}}
\newcommand{\ro}{\boldsymbol{\rho}}
\newcommand{\si}{\boldsymbol{\sigma}}
\newcommand{\thv}{\boldsymbol{\theta}}
\newcommand{\bv}{\boldsymbol{b}}
\newcommand{\JV}{\boldsymbol{J}}
\newcommand{\nv}{\boldsymbol{n}}
\newcommand{\lv}{\boldsymbol{l}}
\newcommand{\om}{\boldsymbol{\omega}}
\newcommand{\Om}{\boldsymbol{\Omega}}
\newcommand{\Piv}{\boldsymbol{\Pi}}
\newcommand{\UV}{\boldsymbol{U}}
\newcommand{\iv}{\boldsymbol{i}}
\newcommand{\nuv}{\boldsymbol{\nu}}
\newcommand{\muv}{\boldsymbol{\mu}}
\newcommand{\lm}{\boldsymbol{\lambda}}
\newcommand{\Lm}{\boldsymbol{\Lambda}}
\newcommand{\opsi}{\overline{\psi}}
\renewcommand{\tan}{\textrm{tg}}
\renewcommand{\cot}{\textrm{ctg}}
\renewcommand{\sinh}{\textrm{sh}}
\renewcommand{\cosh}{\textrm{ch}}
\renewcommand{\tanh}{\textrm{th}}
\renewcommand{\coth}{\textrm{cth}}

\section{Introduction}
Many attempts have been made to extend the relativity of motion in the framework of the special theory of relativity, including transformations between accelerated coordinate systems. While Page \cite{LP} proposed a detailed theory of the kinematical part of the matter, Robertson \cite{HR} has given a penetrating treatment of the general mathematical theory. A link between Page's theory and the 4-dimensional 15 parameters conformal group has been pointed out by Engstrom and Zorn \cite{EZ} whilst Hill \cite{EH} developed a mathematical method (on the basis of Lie theory of the conformal group) on the transformation to accelerated axes which has a geometrical interpretation in terms of the inversive transformation. Although the conformal group contains the 10 parameters  Lorentz group as a subgroup, the physical meaning of some of its operations (dilatation or the special conformal transformation) \cite{FRW} is still under debate.

\section{Special conformal transformation}
 It is a known fact that the special conformal transformation is a part of the 15 parameters conformal group, including the 10 parameters of the Lorentz group. It contains also an extra dilatational parameter and another four characterizing the special conformal transformation  \cite{EH, CCJ, HC1}
 \begin{equation}
 x^{'b} = \frac{x^{b} - a^{b} x_{c}x^{c}}{1 - 2a_{b} x^{b} + a^{2} x_{b} x^{b}}
\label{2.1}
\end{equation}
where $a^{2} = \eta_{bc}a^{b}a^{c}, x^{c}x_{c} = \eta_{ab} x^{a}x^{b}$ is the Minkowski interval, $\eta_{ab} = diag(-1, 1, 1, 1)$ and $a^{b}$ is a constant vector with dimension of acceleration \cite{HC1}. The latin indices run from 0 to 3 (in the order $t, x, y, z$) and we take $G = c = 1$, unless otherwise specified. The coordinate transformation (2.1) becomes the Newtonian acceleration transformation when $c \rightarrow \infty$, which leads to $x' = x + at^{2}$ (for unidimensional uniformly accelerated motion, the classical acceleration being $2a > 0$), with $a^{b} = (0, a, 0, 0)$. It is composed of the following set (inversion - translation - inversion) 
 \begin{equation}
 x^{b} \rightarrow \frac{k^{2} x^{b}}{x^{a}x_{a}} \rightarrow k^{2}(\frac{x^{b}}{x^{a}x_{a}} - a^{b}) \rightarrow x^{'b}
\label{2.2}
\end{equation} 
where the constant $k$ (with units of length) was introduced for dimensional reasons.

The above set of transformation changes the Minkowski line element $ds^{2} = \eta_{ab}dX^{a}dX^{b}, X^{a} = (T, X, Y, Z)$ into 
 \begin{equation}
 ds^{2} = \frac{1}{(1 - 2a_{b} x^{b} + a^{2} x_{b} x^{b})^{2}} \eta_{ab}dx^{a}dx^{b}
\label{2.3}
\end{equation}
 The metric (2.3) may also be written as
 \begin{equation}
 ds^{2} = \frac{-dt^{2} +dx^{2} + dy^{2} +dz^{2}}{[(1 - ax)^{2} - a^{2}(t^{2} - y^{2} - z^{2})]^{2}} 
\label{2.4}
\end{equation}
 The geometry (2.3) is, of course, flat, being obtained from the Minkowski metric by means of the set of transformations (2.2). It is singular on the expanding null sphere $(x - \frac{1}{a})^{2} +y^{2} +z^{2} = t^{2}$, with the center located at $(1/a, 0, 0)$. The spacetime (2.3) is not spherically symmetric due to the special direction of $a$. It acquires spherical symmetry if the translation $x \rightarrow x - 1/a$ is performed.
 
 \section{Near x-axis case}
 Let us now consider that our accelerated observer is located close to the $x$-axis, so that $ \sqrt{y^{2} + z^{2}} << x$. In other words, he finds within a thin tube centered on the $x$-axis. Therefore (2.4) becomes now
 \begin{equation}
 ds^{2} = \frac{-dt^{2} +dx^{2} + dy^{2} +dz^{2}}{[(1 - ax)^{2} - a^{2}t^{2}]^{2}} 
\label{3.1}
\end{equation}
The geometry (3.1) is no longer flat because of the approximation used. We encounter a similar situation near the black hole horizon where the Schwarzschild metric acquires a Rindler form \cite{HC2}
  \begin{equation}
ds^{2} = -\frac{r-2m}{2m} dt^{2} + \frac{2m}{r-2m} dr^{2} + 4m^{2} (d\theta^{2} + sin^{2} \theta d\phi^{2})
\label{3.2}
\end{equation}
where $m$ is the black hole mass. Nevertheless, only the two-dimensional form of (3.2) is flat ($\theta, \phi = const.$ or $\Delta\theta$ and $\Delta\phi$ are negligible w.r.t. $\pi$). The restriction $r \approx 2m$ only is not enough to get flat Rindler spacetime. We may actually travel around the black hole close to the horizon and feel its curvature. That could be seen from the nonzero component of the Riemann tensor for the metric (3.2), namely $R^{\theta \phi}~_{ \theta \phi} = 1/4m^{2}$. 

The spacetime (3.1), not being flat, it would be interesting to find what is the source of curvature, i.e. the stress tensor from the r.h.s. of Einstein's equations $G_{ab} = 8\pi T_{ab}$. The calculations give us the nonzero components
  \begin{equation}
  T^{t}_{~t} = -\rho = \frac{a^{2}}{\pi} [(1 - ax)^{2} - a^{2}t^{2}] = p_{x} = 2p_{\bot}
\label{3.3}
\end{equation}
where $p_{x} = T^{x}_{~x}$ is the pressure of the anisotropic fluid along the $x$-axis and $p_{\bot}$ refers to the transverse coordinates $y$ and $z$. Noting that $T^{a}_{~b}$ is diagonal, so the fluid is comoving. For the time being we consider $f \equiv f(x,t) = (1 - ax)^{2} - a^{2}t^{2} > 0$, which covers two quadrants in $x-t$ plane, bounded by the null surfaces $x = \pm{t} + (1/a)$. These surfaces are null geodesics (obtained from $ds^{2} = 0$) for a photon traveling parallel with the $x$-axis ($y, z$ = const.). From (3.3) we observe that $\rho < 0;~ p_{x},~ p_{\bot} > 0$ and the spacetime has negative curvature: $R^{a}_{~a} = -24a^{2}f < 0$. In addition, the Kretschmann scalar is given by $R_{abcd}R^{abcd} = 128a^{4}f^{2}$. We must stress that, even though the metric (3.1) is singular for $f = 0$,~however $ \rho, p_{x}, p_{\bot}, R^{a}_{~a}$ and the Kretschmann scalar are regular there. Moreover, they vanish at the boundaries of the spacetime, $f = 0$. It is worth noting that, a ''static'' observer in the accelerated system (say, $x = x_{0} =$ const.) moves hyperbolically w.r.t. a Minkowski observer. To show this, we have to eliminate $t$ between the equations
  \begin{equation}
  T = \frac{t} {(1 - ax_{0})^{2} - a^{2}t^{2}},~~~X = \frac{x - a(x^{2} - t^{2})} {(1 - ax_{0})^{2} - a^{2}t^{2}}
\label{3.4}
\end{equation}
One obtains the hyperbola
  \begin{equation}
  \left[X + \frac{1}{a} - \frac{1}{2a(1 - ax_{0})}\right]^{2} - T^{2} = \frac{1}{4a^{2}(1 - ax_{0})^{2}}
\label{3.5}
\end{equation}
Let us now take into consideration a congruence of ''static'' observers with the velocity vector field
  \begin{equation}
  u^{a} = (f, 0, 0, 0),~~~u^{a}u_{a} = -1
\label{3.6}
\end{equation}
in the spacetime (3.1). One may check that the above congruence is not geodesic, the acceleration 4 - vector being given by
  \begin{equation}
   A^{b} \equiv  u^{a} \nabla_{a} u^{b} = (0,2a(1-ax)f,~0,~0), ~~~\sqrt{g_{bc}A^{b} A^{c}} = 2a|1 - ax|,
\label{3.7}
\end{equation}
where $g_{bc}$ corresponds to the metric (3.1). One sees from (3.7) that $A^{x} > 0$ in one of the two quadrants (where $x > 1/a$) and $a^{x} < 0$ in the other quadrant ($x < 1/a$). This could be also noticed from $A^{b}n_{b} = 2a(1 - ax)$, where $n^{b} = (0, f, 0, 0)$ is the normal to the $x = $~const. surface.

 As far as the scalar expansion of the congruence is concerned, (3.6) yields
  \begin{equation}
    \Theta \equiv \nabla_{a} u^{a} = \frac{1}{\sqrt{-g}} \frac{\partial}{\partial x^{a}} (\sqrt{-g} u^{a}) = 6a^{2}t,
\label{3.8}
\end{equation}
where $g$ is the determinant of the metric (3.1). For the time evolution of the expansion we have
  \begin{equation}
   \dot{\Theta} \equiv  u^{a} \nabla_{a}\Theta = 6a^{2}f,  
\label{3.9}
\end{equation}
which is always positive. We observe that $\Theta \propto t$ and $ \dot{\Theta} > 0$ in the chosen quadrants. One results that the anisotropic fluid expands for $t > 0$ and shrinks for $t < 0$. We also obtain that $\sigma_{ab} = 0,~\omega_{ab} = 0$, where $\sigma_{ab}$ and $\omega_{ab}$ are, respectively, the shear and vorticity tensors of the fluid.

The trace of the stress tensor is given by $T^{a}_{~a} = 3a^{2}f/\pi >0$. It is vanishing on the null boundary $f = 0$, i.e. $x = \pm{t} + (1/a)$, as expected. $T_{a}^{~b}$ can be put in a general form, characterizing an anisotropic fluid without heat flux
  \begin{equation}
  T_{~a}^{b} = (\rho + p_{\bot})u_{a} u^{b} + p_{\bot} \delta_{a}^{b}+ (p_{x} - p_{\bot}) n_{a} n^{b},
\label{3.10}
\end{equation}
with $n^{a}u_{a} = 0$. 

It is instructive to investigate the behaviour of our physical system on the other two quadrants of the $x - t$ plane, where $f < 0$. The result is that $R_{ab}, G_{ab}, R^{a}_{~a}, R_{abcd}$ change their sign, preserving their form. Hence, the scalar curvature is now positive and $\rho > 0$, while $p_{x}, p_{\bot} < 0$. In addition, we may no longer have $u^{t} = f$, because $u^{t} = dt/d\tau$ should be positive ($\tau$ is the proper time). Therefore, it is mandatory to take $u^{a} = (-f, 0, 0, 0)$ from the very beginning. With this new $u^{a}$, the scalar expansion $\Theta = -6a^{2}t$ and  $\dot{\Theta} = 6a^{2}f$ change their sign.  

\section{Static approximation}
Let us focus now on the static situation. We shall therefore neglect the second order terms containing $a^{2}x^{2}$ and $a^{2}t^{2}$ in the metric (3.1), apart from the smallness of $y$ and $z$,  preserving the first order terms in acceleration (our observer is considered to be located close to the origin of coordinates in the plane $x - t$). Consequently, the geometry (3.1) acquires the static conformally flat form
 \begin{equation}
 ds^{2} = \frac{-dt^{2} +dx^{2} + dy^{2} +dz^{2}}{(1 -2ax)^{2}}  
\label{4.1}
\end{equation}
Our approximation resembles the Majhi and Padmanabhan (MP) one \cite{MP}. They retained only the terms linear in the surface gravity $\kappa$ when they computed the infinitesimal form of the coordinate transformation leading to the Rindler and ''near horizon'' Schwarzschild (in Kruskal coordinates) metrics, respectively (see their Eqs. 28 and A3). The transformation (2.1) becomes in the static approximation
 \begin{equation}
   T = \frac{t} {1 - 2ax} \approx t(1 + 2ax),~~~X = \frac{x - a(x^{2} - t^{2})} {1 - 2ax} \approx x +a(x^{2} + t^{2})
\label{4.2}
\end{equation}
where $(T, X)$ are the Minkowski coordinates. 

It is a well-known that the set of transformations (2.1) is reversible, namely it preserves the final form (when we pass from $x'^{c}$ to $x^{c}$, with the only change $a^{b} \rightarrow -a^{b}$), as for the Lorentz transformation, where we have $v \rightarrow -v$. One observes from (4.2) that the generators of the infinitesimal transformation 
 \begin{equation}
 q^{t} \equiv t - T = -2axt,~~~q^{x} \equiv x - X = -at^{2} - ax^{2}
\label{4.3}
\end{equation}
are similar with those obtained by MP \cite{MP}(Eq. A3, where their tortoise coordinate $r_{*}$ corresponds to our $x$ and the surface gravity $\kappa$ corresponds to $2a$, as it should be). The fact that the infinitesimal transformation concide for the special conformal transformation and for the Kruskal-like metric from the MP paper deserves, in our opinion, a deeper analysis.

We wish now to investigate what sources are necessary on the r.h.s. of Einstein's equations for the geometry (4.1) to be an exact solution. A straightforward calculation gives us
 \begin{equation}
   T^{t}_{~t} = -\rho = \frac{3a^{2}}{2\pi},~~~p = T_{~x}^{x} = T_{~y}^{y} = T_{~z}^{z} = \frac{3a^{2}}{2\pi}
\label{4.4}
\end{equation}
In other words, this time we deal with a perfect fluid given by the equation of state $p = -\rho$. Hence, the energy-momentum tensor represents a negative ''cosmological constant'' $\Lambda = -12a^{2}$ (noting that the scalar curvature is also negative, $R^{a}_{~a} = -48a^{2}$, but the Kretschmann scalar acquires the value $K = 384a^{4}$). Surprisingly, the stress tensor and the curvature invariants are constant everywhere, even though the metric is $x - $ dependent. Moreover, the energy density is proportional with the acceleration squared, as in Newtonian mechanics. As an order of magnitude, we evaluate the energy density for $a = 10^{3} cm/s^{2}$. One obtains $\rho = -3a^{2}/2\pi G \approx -10^{15} erg/cm^{3}$. For the ''cosmological constant'' we have $|\Lambda| = 12a^{2}/c^{4} \approx 5.10^{-36} cm^{-2}$. 

Taking again a congruence of static observers with the velocity field $u^{a} = (1 - 2ax, 0, 0, 0)$, one obtains for the kinematical parameters of the congruence
 \begin{equation}
 A^{b} = (0, 2a(1 - 2ax), 0, 0),~~~\sqrt{g_{bc}A^{b} A^{c}} = 2a,~~~\Theta = 0, ~~~\sigma_{ab} = \omega_{ab} = 0.
\label{4.5}
\end{equation}
We get now a constant invariant acceleration. The nongeodesic congruence has a vanishing expansion due to the static character of the metric used. Since $A^{x} = 2a(1 - 2ax) > 0$, the field is attractive. Though the metric (4.1), being curved, cannot be obtained from Minkowski's geometry by a coordinate transformation, the interpretation of $2a$ as the invariant acceleration is still valid.

\section{Conclusions}
 We generate a time dependent conformally flat spacetime by means of the special conformal transformation of the 15 parameters conformal group that contains the Lorentz group as a subgroup. Neglecting the variation of the metric with the transversal coordinates, the new spacetime obtained is an exact solution of the Einstein equations with a stress tensor of an anisotropic fluid with negative energy density. The boundaries of the spacetime are the null surfaces $x = \pm{t} + (1/a)$. All the curvature invariants are finite in the entire spacetime. In addition, the stress tensor corresponding to the static approximation is of $\Lambda$ - form, with an energy density proportional to the acceleration squared, as in Newtonian mechanics.

\end{document}